\documentclass[twocolumn,nofootinbib,showpacs,preprintnumbers,amsmath,amssymb]{revtex4}

\usepackage{graphicx}
\usepackage{dcolumn}
\usepackage{bm}
\newcommand{\be}{\begin{equation}}
\newcommand{\ee}{\end{equation}}
\newcommand{\bee}{\begin{eqnarray}}
\newcommand{\eee}{\end{eqnarray}}

\begin{document}
\title{Momentum distributions, spin-dependent observables,
and the $\boldsymbol{D_2}$ parameter \\
for $^3$He breakup}
\author{A.~P.~Kobushkin}
 \email{kobushkin@bitp.kiev.ua}
\affiliation{%
Bogolyubov Institute for Theoretical Physics, Metrologicheskaya str.\ 14B,
03680 Kiev, Ukraine
}%
\affiliation{National Technical University of Ukraine KPI,
Prospekt Peremogy 37, 03056 Kiev, Ukraine}
\author{E.~A.~Strokovsky}
 \email{Eugene.Strokovsky@sunse.jinr.ru}
\affiliation{%
Laboratory of High Energy Physics, Joint Institute for Nuclear Research, 141980, Dubna, Russia
}%

\date{\today}

\begin{abstract}
Using a recent parametrization of the fully antisymmetric three-nucleon wave
function, based on the Paris and CD-Bonn potentials, we analyze the momentum
distributions of constituents in $^3$He as well as the spin-dependent
observables for $(^3\mathrm{He},d)$ and $(^3\mathrm{He},p)$ breakup
reactions. Special attention is paid to the region of small relative momenta of
the $^3$He constituents, where a single parameter, $D_2$, has a determining role for
the spin-dependent observables in the $d+p$ channel. This fact results in some useful
relations between experimental observables.
\end{abstract}

\pacs{25.30.Bf,13.40.-f,13.60.-Hb,13.88.+e}
\maketitle

\section{\label{sec:Introduction}Introduction}

Momentum distributions of one and two nucleon fragments in $^3$He give
important information about nuclear systems more complicated than the
deuteron. They cast light on such interesting problems as the nucleon-nucleon
interaction at short distances, the role of three-body interaction (the
3{\it N} forces), and non-nucleonic degrees of freedom in nuclei.

Precise data are currently available on the momentum distributions of the
proton and the deuteron obtained with
electromagnetic~\cite{em_probe1,em_probe2,em_probe3,em_probe4} and hadronic
probes~\cite{Ableev3He,Kitching,Epstein}. Data on the energy dependence of
the differential cross sections of backward elastic
$^3\text{He}(p,^{3\!}\text{He})p$ scattering, which are related to the same
momentum distributions, also exist~\cite{3hebes,RCNP}. Furthermore, the
spin-correlation parameter $C_{yy}$ for this reaction was recently measured
for first time~\cite{RCNP}. Finally, the tensor polarization of the deuteron
(sometimes called its ``alignment'')
in the $^{12}\mathrm{C}(^3\mathrm{He},d)$ reaction was also measured~\cite{Sitnik,
Sitnik_prel}. Both this and the $C_{yy}$ data~\cite{RCNP} are sensitive to
the spin structure of $^3$He.

For a long time the lack of useful parametrizations of the three-nucleon wave
function has made the theoretical analysis of these data difficult. However,
a convenient parametrization of the fully antisymmetric three-nucleon wave
function based on the Paris~\cite{Paris} and CD-Bonn~\cite{CD-Bonn}
potentials has recently been presented~\cite{Baru}, which we use here
to calculate the momentum distributions in $^3$He, as well
as the spin-dependent observables, within the framework of the spectator
model for the $^3$He breakup reactions. Special attention is paid to the study of the
two-body $^3\mathrm{He}\rightarrow d+p$ channel. We compare our results with other
theoretical works and with existing experimental data.

In our analysis of spin-dependent observables for $(^3\mathrm{He},d)$ and
$(^3\mathrm{He},p)$ reactions, we carefully consider their behavior in the
region of small (below $\sim 150$~MeV/$c$) internal momenta of the $^3$He
fragments, where a single quantity, known in the literature as the $D_2$
parameter, completely determines both the sign and the momentum dependence of
the observables. Several relations between experimental observables in this
region were obtained from this consideration. These provide useful checks on
the consistency of the experimental database of spin-dependent observables
because at low relative momenta of the $^3$He fragments the model used for
the breakup reactions works reasonably well.

The paper is organized as follows. In Sec.~\ref{Sec:2} we discuss the
parametrization of the three-nucleon wave function used in the later work.
Distributions of one- and two-nucleon constituents in $^3$He are evaluated
and compared with the results of other calculations in Sec.~\ref{sec:MD}.
Various spin-dependent observables for $(^3\mathrm{He},d)$ and
$(^3\mathrm{He},p)$ breakup are considered and discussed in
Sec.~\ref{Sec:Spin}. In Sec.~\ref{Sec:Comparing} we compare our results with
experiment within the ``minimal relativization scheme'' and discuss briefly
the limitations of our approach. Conclusions and discussion are presented in
Sec.~\ref{Sec:Conclusions}.

\begin{table*}
\caption{\label{tab:QN}Quantum numbers of the $^3\mathrm{He}$ partial waves. Here
$s$, $\tau$, $\ell$, and $j$ are spin, isospin, orbital, and total
angular momenta of the pair; $L$ and $K$ are relative angular momenta for the
spectator and the channel spin, respectively.}
\begin{ruledtabular}
\begin{tabular}{cccccccc}
Channel No.  &  Label    &   $\ell$ & $s$& $j^\pi$ & $K$ & $L$ & $\tau$ \\
\hline
1           &  $\qquad ^1s_0S \qquad $ &   $\qquad 0\qquad $     & $\qquad 0 \qquad $ & $\qquad 0^+ \qquad $   & $\qquad1/2\qquad $ & $\qquad 0\qquad $  &  $\qquad 1\qquad $     \\
2           &  $^3s_1S$ &    0     & 1  & $1^+$   & 1/2 &  0  &  0     \\
3           &  $^3s_1D$ &    0     & 1  & $1^+$   & 3/2 &  2  &  0     \\
4           &  $^3d_1S$ &    2     & 1  & $1^+$   & 1/2 &  0  &  0     \\
5           &  $^3d_1D$ &    2     & 1  & $1^+$   & 3/2 &  2  &  0     \\
\end{tabular}
\end{ruledtabular}
\end{table*}

%
\section{\label{Sec:2}The parametrization of the three-nucleon wave function}

We here give a review of the parametrization of the $\mathrm{^3He}$ wave
function~\cite{Baru}. Working in the framework of the so-called channel spin
coupling scheme (Ref.~\cite{Schadow}), the authors of Ref.~\cite{Baru}
restricted themselves to five partial waves,
\be
\label{Channelspin}
\left|\left\{\left[(\ell s)j\tfrac12\right]KL\right\}\tfrac12\right>,
\ee
where $\ell$, $j$, and  $s$ are the orbital, total, and spin angular momenta
for the pair (the second and third nucleons), and $L$ and $K$ are the
relative orbital angular momentum for the spectator (the first
nucleon) and the channel spin, respectively. Coulomb effects are not
included. The appropriate quantum numbers of the partial waves are collected
in Table~\ref{tab:QN}.

We use the standard definition of the Jacobi coordinates $\mathbf{r}$ and
\mbox{\boldmath$\rho$} in the three-particle system and the corresponding
momenta $\mathbf{p}$ and $\mathbf{q}$:%
\be\label{Jacobi}
\begin{array}{ll}
\mathbf{r}_1=\mathbf{R}+\tfrac23\mbox{\boldmath$\rho$}, & \mathbf{p}_1=\tfrac13\mathbf{P} + \mathbf{q}\,,\\[0.25cm]
\mathbf{r}_2=\mathbf{R}-\tfrac13\mbox{\boldmath$\rho$}+\tfrac12 \mathbf{r},\ \ & \mathbf{p}_2=\tfrac13\mathbf{P} - \tfrac12\mathbf{q} + \mathbf{p}\,,\\[0.25cm]
\mathbf{r}_3=\mathbf{R}-\tfrac13\mbox{\boldmath$\rho$}-\tfrac12 \mathbf{r},\  & \mathbf{p}_3=\tfrac13\mathbf{P} - \tfrac12\mathbf{q} - \mathbf{p}\,.\\[0.25cm]
\end{array}
\ee
Here $\mathbf{R}$ is the coordinate of the nucleus center of mass (with
$\mathbf{P}$ being the total momentum of the nucleus), \mbox{\boldmath$\rho$}
is the radius vector from the center of mass of the nucleon pair to the
nucleon 1 (the corresponding momentum is $\mathbf{q}$), and $\mathbf{r}$ is
the separation between the nucleons in the pair (the corresponding momentum
is $\mathbf{p}$).

\begin{figure*}[t]
  \centering
   \includegraphics[height=0.35\textheight]{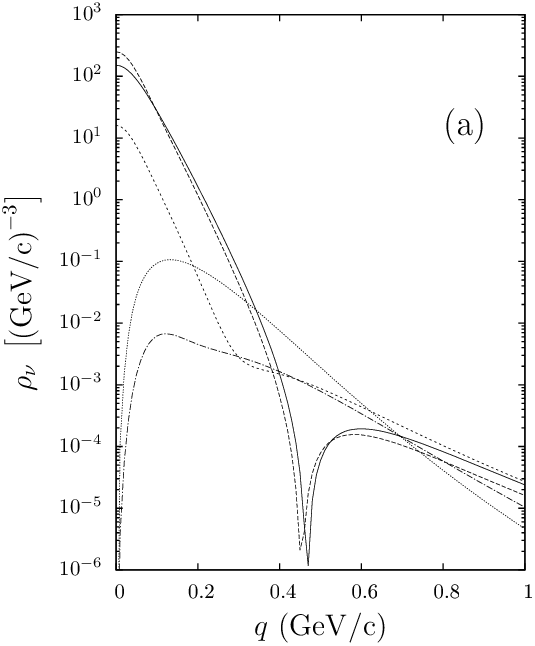}\hspace{1.cm}
   \includegraphics[height=0.35\textheight]{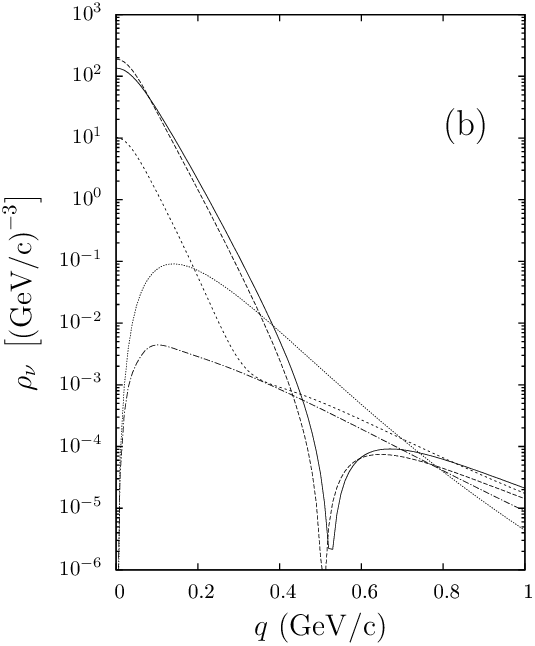}
\caption{One-nucleon momentum distributions in partial channels for the
Paris~\cite{Paris} (a) and CD-Bonn~\cite{CD-Bonn} (b) potentials. Long-dashed
lines, $^1s_0S$; solid lines, $^3s_1S$; short-dashed lines, $^3d_1S$; dotted
lines, $^3s_1D$; and dash-dotted lines, $^3d_1D$. } \label{fig:1}
\end{figure*}

Explicitly, the wave function of $\mathrm{^3He}$ in momentum space reads
\begin{widetext}
\begin{equation}
\label{Explicity}
\begin{split}
\Psi_\sigma(\mathbf p, \mathbf q\,)=
&\sum_\xi\left\{\frac{1}{4\pi}\delta_{\xi\sigma}
\sum_{\tau_3,t_3}
\left<1\tfrac12 \tau_3t_3\bigm|\tfrac12\tfrac12\right>\psi_1(p,q)
\left|00;1\tau_3\right>\chi_{\xi t_3}+
\sum_{s_3}\left[
\frac{1}{4\pi}\left<1\tfrac12s_3\xi\bigm|\tfrac12\sigma\right> \psi_2(p,q) \right.\right.\\
&\left.\left. -\sqrt{\frac{1}{4\pi}}\sum_{L_3K_3}\left<1\tfrac12s_3\xi\bigm|\tfrac32 K_3\right>
\left<\tfrac32 2 K_3 L_3 \bigm|\tfrac12\sigma\right> Y_{2L_3}(\widehat {\mathbf q}\,)\psi_3(p,q) \right.\right.\\
&\left.\left. -\sqrt{\frac{1}{4\pi}}\sum_{\ell_3M}\left<12 s_3\ell_3|1M\right>
\left<1\tfrac12M\xi\bigm|\tfrac12 \sigma\right>  Y_{2\ell_3}(\widehat {\mathbf p}\,)\psi_4(p,q)\right.\right.\\
&\left.\left.
+\sum_{\ell_3ML_3K_3}\left<12 s_3\ell_3|1M\right>
\left<1\tfrac12M\xi\bigm|\tfrac32 K_3\right>
\left<\tfrac32 2 K_3 L_3\bigm|\tfrac12\sigma \right> 
Y_{2L_3}(\widehat {\mathbf q}\,) Y_{2\ell_3}(\widehat {\mathbf p}\,)\psi_5(p,q)
\right] \left|1s_3;00\right>\chi_{\xi\frac12}\right\}\  {,}
\end{split}
\end{equation}
\end{widetext}
where $\sigma$ and $\xi$ are the spin projections of $^3$He and the
spectator nucleon, $t_3$ is the isospin projection of the spectator nucleon,
$M$ is the projection of the total angular momentum of the pair, and
$\chi_{\xi t_3}$ and $\left|ss_3;\tau\tau_3\right>$ are the spin-isospin wave
functions of the spectator nucleon and the pair, respectively.

Note, that in Eq.~(\ref{Explicity}) we use the following convention
for angular momentum summation
\be\label{convention}
j+\tfrac12\to K,\qquad K+L \to \tfrac12.
\ee
Other conventions are often used in the literature, for example:
\be\label{alternative_convention}
j+\tfrac12\to K,\qquad L+K \to \tfrac12
\ee
and
\be\label{friar convention}
\tfrac12+j\to K,\qquad L+K \to \tfrac12.
\ee
The convention of Eq.~(\ref{alternative_convention}) was used, in particular,
in Ref.~\cite{Knutson}, whereas that of Eq.~(\ref{friar convention}) was
exploited in Ref.~\cite{friar}.

Due to the properties of the Clebsch-Gordan coefficients under permutations,
for example $\left< \tfrac32\, 2\, K_3\, L_3 | \tfrac12 \xi \right>=-\left<
2\, \tfrac32\, L_3\, K_3 | \tfrac12 \xi \right>$, some of the wave function
components have opposite signs in different conventions. For example, using
Eq.~(\ref{alternative_convention}) rather than Eq.~(\ref{convention}) would
result in $\psi_3(p,q) \to -\psi_3(p,q)$ and $\psi_5(p,q) \to -\psi_5(p,q)$.
Similarly, the use of Eq.~(\ref{friar convention}) instead of
Eq.~(\ref{convention}) would give $\psi_2(p,q) \to -\psi_2(p,q)$,
$\psi_3(p,q) \to -\psi_3(p,q)$, $\psi_4(p,q) \to -\psi_4(p,q)$, and
$\psi_5(p,q) \to -\psi_5(p,q)$, while $\psi_1(p,q)$ would not change sign.

The partial waves are approximated by the following functions of the two
momenta $p$ and $q$:
\begin{equation}
\psi_\nu(p,q)=v^\nu_1(p)w^\nu_1(q) +v^\nu_2(p)w^\nu_2(q),
\label{eq:expan}
\end{equation}
where
\be
\begin{split}
& v^\nu_i(p)=\sum_{n=1}^5\frac{a_{n,i}^\nu}{p^2 + (m^\nu_{n,i})^2},
\\
& w^\nu_i(q)=\sum_{n=1}^5\frac{b_{n,i}^\nu}{q^2 + (M^\nu_{n,i})^2}.
\end{split}
\label{eq:vw} \ee
The parameters are restricted by the usual constraints for
$S$ and $D$ waves:
\bee \sum_{n=1}^5a_{n,i}^\nu = \sum_{n=1}^5b_{n,i}^\nu=0.
\label{eq:constarSD} \eee There are additional constraints for $D$ waves: \be
\begin{split}
&\sum_{n=1}^5a_{n,i}^\nu(m^\nu_{n,i})^2 = \sum_{n=1}^5b_{n,i}^\nu(M^\nu_{n,i})^2=0,
\\
&\sum_{n=1}^5\frac{a_{n,i}^\nu}{(m^\nu_{n,i})^2} =
\sum_{n=1}^5\frac{b_{n,i}^\nu}{(M^\nu_{n,i})^2}=0.
\end{split}
\label{eq:constarD} \ee
The full $\mathrm{^3He}$ wave function is normalized by the condition
 \be \sum_{\nu=1}^5\int_0^\infty dq\int_0^\infty dp\,
 q^2p^2\,|\psi_\nu(p,q)|^2=1. \label{eq:Norm} \ee
\section{\label{sec:MD}Momentum distributions}
\subsection{\label{sec:OneNucl}One-nucleon distributions}

Assuming final plane waves, the  one-nucleon momentum distribution
in a partial channel $\nu$ becomes
 \bee \rho_{\nu}(q)=\frac3{4\pi}\int_0^\infty dp\, p^2 \left|
 \psi_{\nu}(p,q)\right|^2, \label{MD_of_pair} \eee
where the coefficient of 3 is the square of a spectroscopic factor. The
distributions for each partial channel are displayed in Fig.~\ref{fig:1}. It
is important to note that the distributions for the $^1s_0S$ and $^3s_1S$
channels are very similar in both their magnitude and their
momentum dependence.

The values of the partial channel probabilities, defined as
\be\label{PW_prob} P_\nu=\frac{1}{3}\int d^3q\,\rho_{\nu}(q)= \int
dp\,dq\,p^2q^2|\psi_\nu(p,q)|^2, \ee
are given in Table~\ref{tab:Probability}.
\begin{figure*}[t]
  \centering
  \includegraphics[height=0.35\textheight]{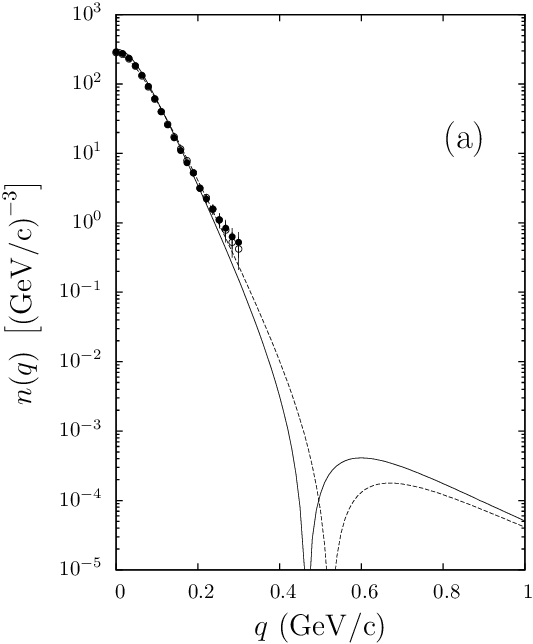}\hspace{1.cm}
  \includegraphics[height=0.35\textheight]{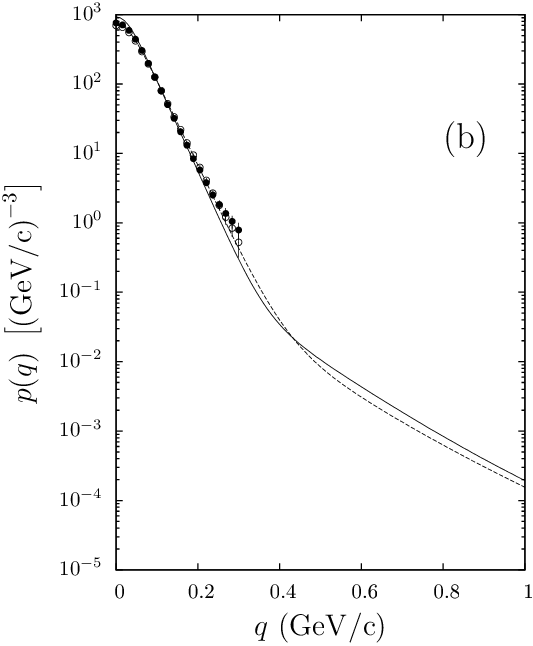}
\caption{The neutron (a) and proton (b) momentum distributions. The solid and
dashed curves are for the Paris and CD-Bonn potentials, respectively. The
solid and open circles represent results of variational
calculations~\cite{Schiavilla} obtained using the Urbana and Argonne
potentials, respectively. } \label{fig:neutron}
\end{figure*}

The momentum distribution of a nucleon $N$ with spin and isospin projections
$\xi$ and $t_3$ in $^3$He with spin projection $\sigma$ is
\be
\label{OneNuclMD}
N_{\sigma (\xi t_3)}(\mathbf q) =
3\sum_{ss_3\tau\tau_3}\int d^3p\left|
\chi^\dag_{\xi t_3}\left<ss_3\tau\tau_3\right|\Psi_\sigma (\mathbf p,\mathbf q)\right|^2.
\ee

In the neutron case, Eq.~\eqref{OneNuclMD} reduces to~\footnote{It must be
noted that normalizations of the proton ($N_p$) and neutron ($N_n$) momentum
distributions in Table 7 of Ref.~\cite{Schiavilla} differ from those used
here; they can be compared with our $n(q)$ and $p(q)$ after multiplication by
a renormalization factor of $(2\pi)^{-3}$.}
\be\label{NeuteronMD}
\begin{split}
n_{\sigma \xi}(q)&=\delta_{\sigma \xi}\frac1{2\pi}\int_0^\infty %
\left|\psi_{1}(p,q)\right|^2p^2dp\\
&=\tfrac23\delta_{\sigma \xi}\rho_1(q) \equiv \delta_{\sigma \xi}n(q).
\end{split}
\ee

Note that the number of neutrons $\mathcal N_n$ in $^3$He is given by the
integral $\mathcal N_n=\int d^3q\,n(q)=1$ and that the $\psi_{1}$ component
must be normalized as
\be\label{NeuteronNorm} \int dp\, dq\,
p^2q^2 \left|\psi_{1}(p,q)\right|^2=\textstyle\frac12\;. \ee
Here and below we use $p_{\sigma \xi}$ and $n_{\sigma \xi}$ instead of
$N_{\sigma (\xi, \frac12)}$ and $N_{\sigma (\xi, -\frac12)}$, respectively.

The resulting neutron momentum distributions, $n(q)$, for the Paris and
CD-Bonn potentials are shown in Fig.~\ref{fig:neutron}(a). We also compare
the results with variational calculations~\cite{Schiavilla} based on the
Argonne and Urbana potentials.


\begin{table*}
\caption{\label{tab:Probability}The partial channel probabilities
$P_{\nu}$ in $^3\mathrm{He}$.}
\begin{ruledtabular}
\begin{tabular}{lccccc}
           &  $^1s_0S$   &$^3s_1S$  & $^3s_1D$ &  $^3d_1S$ &$^3d_1D$ \\
\hline
Paris      & 0.5000         & 0.4600  & 0.0282 & 0.0103   & 0.0015 \\
CD-Bonn    & 0.5000         & 0.4658  & 0.0231 & 0.0102   & 0.0009\\
\end{tabular}
\end{ruledtabular}
\end{table*}
\begin{figure*}
  \centering
  \includegraphics[height=0.35\textheight]{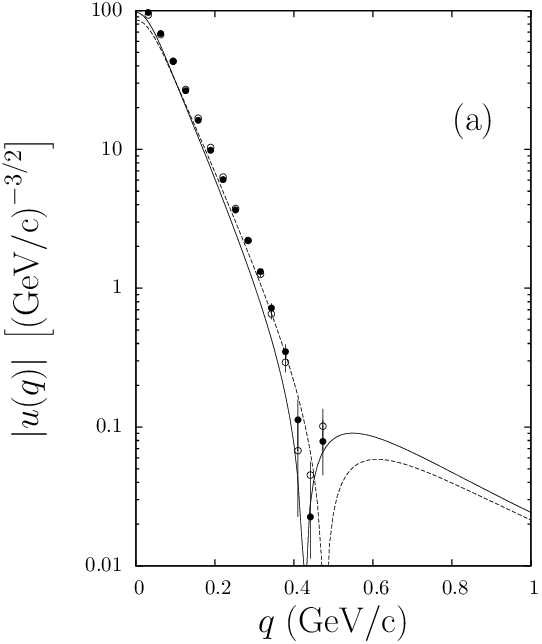}\hspace{1.cm}
  \includegraphics[height=0.35\textheight]{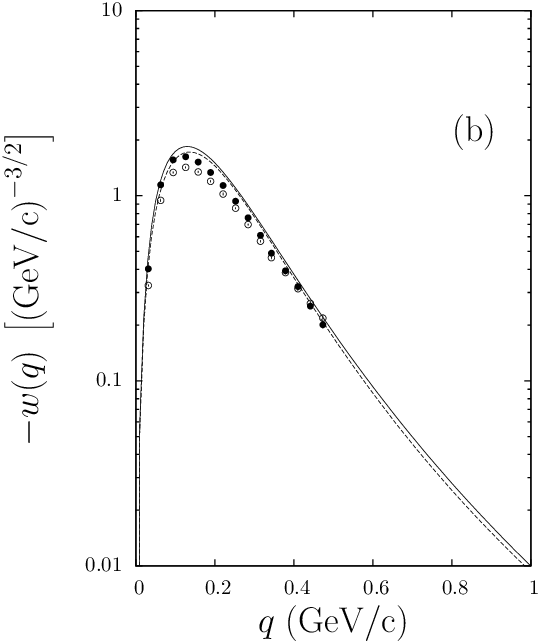}
\caption{The $u(q)$ (a) and $w(q)$ (b) wave functions of the relative
deuteron-proton motion in $^3$He. The notations are the same as those
in Fig.~\ref{fig:neutron}.} \label{fig:u_and_w}
\end{figure*}

Using the results given in Appendix, we get the following for the momentum distribution
of protons with spin projection $\frac12$ and $-\frac12$ in the $^3$He having spin
projection~$\mathbf{+}\frac12$:
\be
\begin{split}
&p_{\frac12\frac12}(q,\theta)=
\frac{1}{3}\left[\rho_{1}(q)+\rho_{2}(q)+\rho_{4}(q)\right] \\\
&\hspace{0.25cm}+\frac{1}{2}\left(\cos^2\theta+\frac13\right)[\rho_{3}(q)+\rho_{5}(q)]  \\
& \hspace{0.25cm}+\sqrt2\left(\frac{1}{3}-\cos^2\theta\right) \left[ \rho_{23}(q) + \rho_{45}(q)\right]
,\\
&p_{\frac12-\frac12}(q,\theta)
=\frac{2}{3}\left[\rho_{2}(q)+\rho_{4}(q)\right] \\
&\hspace{0.25cm}+\frac{1}{2}\left(-\cos^2\theta+\frac53\right)\left[\rho_{3}(q)+\rho_{5}(q)\right] \\
&\hspace{0.25cm} -\sqrt2\left(\frac{1}{3}-\cos^2\theta\right) \left[ \rho_{23}(q) + \rho_{45}(q)\right],
\end{split}
\label{p_1212}
\ee
where $\theta$ is the angle between the $z$ axis (the quantization axis) and
the proton momentum $\mathbf q$. The function $\rho_{\mu\nu}(q)$ is defined
as
\be\label{Nij}
\rho_{\mu\nu}(q)=\frac3{4\pi}\int_0^\infty dp\,p^2
\psi_\mu(p,q)\psi_\nu(p,q),\qquad \mu\neq \nu. \ee
(When $\mu=\nu$ it is sufficient to retain a single index.)

The momentum distribution of the proton, given by the sum of
$p_{\frac12\frac12}(q,\theta)$ and $p_{\frac12-\frac12}(q,\theta)$, does not
depend upon the angle $\theta$:
\be
\label{pMD}
\begin{split}
p(q)=&p_{\frac12\frac12}(q,\theta)+p_{\frac12-\frac12}(q,\theta)=
\frac{1}{3}\rho_{1}(q)+\rho_{2}(q) \\
&+\rho_{3}(q)+\rho_{4}(q)+\rho_{5}(q).
\end{split}
\ee
Figure~\ref{fig:neutron}(b) displays the proton momentum distribution
calculated using four potentials.

From Eqs.~(\ref{NeuteronNorm}), (\ref{p_1212}), and (\ref{pMD}) and the
normalization condition of Eq.~(\ref{eq:Norm}), it follows immediately that
the number of protons in $^3$He is $\mathcal N_p=\int d^3q\, p(q) =2$.

\subsection{Two-nucleon momentum distributions}

We define the two-body amplitudes $A_{dp}(M,\xi,\sigma,\mathbf q)$ as
\be\label{d-p}
\begin{split}
&A_{dp}(M,\xi,\sigma,\mathbf q)\\
&=\sqrt{3}\int d^3r\, d^3\rho\, \Psi^\dag_d(M,\mathbf
r)\chi^\dag_{\xi\frac12}\exp\left(-i\mbox{\boldmath$\rho$}\cdot\mathbf{q}\right)
\Psi_\sigma(\mathbf{r},
\mbox{\boldmath$\rho$})\\
&=(2\pi)^{\frac32}\sqrt 3\int d^3p\, {\psi_{d}}^\dag(M,\mathbf p)\chi^\dag_{\xi\frac12}\Psi_\sigma(\mathbf p,\mathbf q)\\
&=
(2\pi)^{\frac32}\left\lbrace \sqrt{\frac{1}{4\pi}}\langle 1\tfrac12 M \xi |\tfrac12 \sigma \rangle u(q) \right.\\
&-\left. \sum_{K_3L_3}\langle 1\tfrac12 M \xi|\tfrac32 K_3 \rangle \langle 2
\tfrac32 L_3 K_3 |\tfrac12 \sigma\rangle Y_{2L_3}(\widehat
q)w(q)\right\rbrace ,
\end{split}
\ee
where $\sqrt 3$ is the spectroscopic factor, $\Psi_d(M,\mathbf r)$ and
$\psi_{d}(M,\mathbf p)$ are the deuteron wave function in coordinate and
momentum space, respectively, $M$ and $\xi$ are spin projections of the
deuteron and the proton and
\be
\begin{split}
u(q)&=\sqrt{3}\int_0^\infty dp\, p^2 \left[
u_d(p)\psi_{2}
(p,q) + w_d(p)\psi_{4}
(p,q)
\right],\\
w(q)&=-\sqrt{3}\int_0^\infty dp\, p^2 \left[
u_d(p)\psi_{3}
(p,q)  +
 w_d(p)\psi_{5}
(p,q) \right],
\end{split}
\label{WFD}
\ee
where $u_d(p)$ and $w_d(p)$ are the deuteron $S$ and $D$ wave functions,
respectively~\footnote{ For the convention given by
Eq.~(\ref{alternative_convention}) one must replace $w(q)$ by $-w(q)$. This notation
was used, e.g., in Ref.~\cite{Kobushkin_Deuteron-93}.}.

\begin{figure}
  \centering
 \includegraphics[height=0.35\textheight]{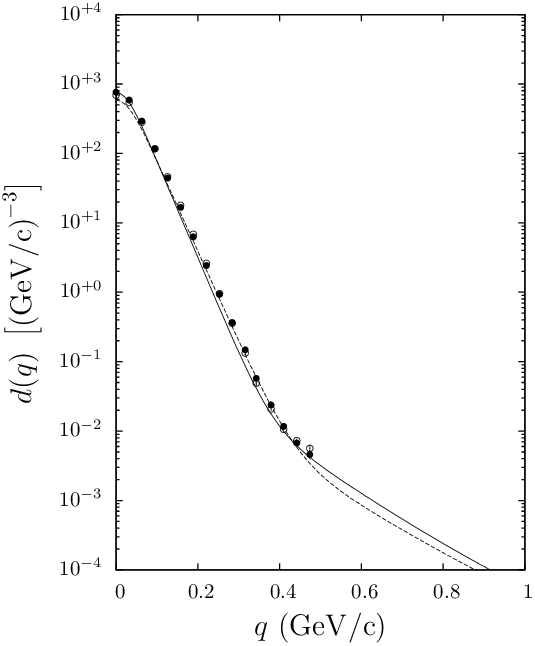}
\caption{The momentum distribution of the deuterons in $^3$He. The notations
are the same as those in Fig.~\ref{fig:neutron}.}
\label{fig:deutMD}
\end{figure}

Equation~(\ref{d-p}) allows us to relate the amplitudes
$A_{dp}^{00}(1,-\tfrac12,\tfrac12,q)$ and
$A_{dp}^{22}(1,-\tfrac12,\tfrac12,q)$ of Schiavilla {\it et al.}~\cite{Schiavilla}
to the wave functions $u(q)$ and $w(q)$:
\be\label{u_and_w}
\begin{split}
u(q)&=\left(\frac1{2\pi}\right)^{\frac32}
\frac{A_{dp}^{00}(1,-\tfrac12,\tfrac12,q)}{\left< 1\frac121-\tfrac12|\tfrac12\tfrac12\right>}\\
&=\left(\frac1{2\pi}\right)^{\frac32}\sqrt{\frac{3}{2}}A_{dp}^{00}(1,-\tfrac12,\tfrac12,q),\\
w(q)&=\left(\frac1{2\pi}\right)^{\frac32}
\frac{A_{dp}^{22}(-1,-\tfrac12,\tfrac12,q)}{\left< 1\frac12-1-\tfrac12|\tfrac32-\tfrac32\right>
\left<\frac32 2 -\tfrac32 2|\tfrac12\tfrac12\right>}\\
&=\left(\frac1{2\pi}\right)^{\frac32}\sqrt{\frac{5}{2}}A_{dp}^{22}(-1,-\tfrac12,\tfrac12,q).
\end{split}
\ee
The $u(q)$ and $w(q)$ wave functions calculated from different potentials are
displayed in Fig.~\ref{fig:u_and_w} and the momentum distribution of the
deuterons,
\be
d(q)=\frac1{4\pi}\left[u^2(q) + w^2(q)\right]
\label{MD_deut}
\ee
is shown in Fig.~\ref{fig:deutMD}.

The effective numbers of the deuterons in $^3\mathrm{He}$, $\mathcal N_d=\int
d^3q\, q^2 d(q)$, are 1.39 and 1.36 for the Paris and CD-Bonn
potentials, respectively.
These can be compared with $\mathcal N_d=1.38$ obtained in variational
calculations~\cite{Schiavilla} with both the Argonne and Urbana potentials.
The probabilities of the $D$ wave in the $d+p$ configuration, $P_D=\int dq\,
q^2 w^2(q)/[\int dq\, q^2(u^2(q)+ w^2(q))]$, are 1.53\% and 1.43\% for the
Paris and CD-Bonn potentials, respectively.

\section{\label{Sec:Spin}Spin-dependent observables}
\subsection{Tensor analyzing powers and the $\boldsymbol{D_2}$ parameter}
In a plane-wave Born approximation the tensor analyzing powers
$T_{20}$, $T_{21}$, and $T_{22}$ of the $(d,t)$ and $(d,^3$He) reactions at
low energies are determined by a single parameter, $D_2$, used, for example, in
Refs.~\cite{Knutson,BH,Roman,Sen}:
\be\label{D2} D_2=\frac1{15}\frac{\int_0^\infty dr\, r^4\,
U(r)}{\int_0^\infty dr\, r^2\, W(r)}=\lim_{q\to 0}\frac{w(q)}{q^2u(q)}, \ee
i.e., $w(q)/u(q)\approx q^2D_2$ at small $q$. In Eq.~(\ref{D2}), $U(r)$
and $W(r)$ are the $S$ and $D$ waves of the $d+p$ component of the $^3$He
wave function in configuration space. The $D_2$ parameter is closely related
to the asymptotic $D$ to $S$ ratio for the $p+d$ partition
of the $^3$He wave function, as is noted in Ref.~\cite{Sen}.

The spin-dependent observables considered here depend upon the bilinear forms
of $S$ and $D$ waves of the $^3$He wave function and the behavior of their
ratio at small $q$ is completely governed by the $D_2$ parameter. In
Table~\ref{tab:D2} we compare this parameter, calculated for the 
wave functions based on different potentials, with the values derived from
experiment.

\subsection{Tensor polarization of the deuteron}
We start by considering the tensor polarization of the deuteron in
($^3$He,$d$) breakup, which is defined as
\be\label{rho20.1}
\rho_{20}=\frac{1}{\sqrt{2}}\frac{d\sigma_+ + d\sigma_- -2d\sigma_0}{d\sigma_+ + d\sigma_- +d\sigma_0}\;,
\ee
where $d\sigma_+$, $d\sigma_-$, and $d\sigma_0$ are the breakup differential
cross sections for the deuteron with spin projections $+1$, $-1$, and $0$
onto the quantization axis, which we take to be along the deuteron momentum,
i.e., $\mathbf q=(0,0,q)$.
In the spectator model the differential cross section
of $^3$He fragmentation in the deuteron with magnetic number $M$ is
proportional to
\be\label{sigj3}
d\sigma(M)\propto\sum_{\xi,\sigma}\left|A_{dp}(M,\xi,\sigma, q\,)\right|^2, \ee
times a coefficient that is independent of $M$.

With this quantization axis, $Y_{2L_3}(\widehat q)\sim \delta_{L_30}$, and
the cross sections are proportional to
\be\label{xsection.2}
\begin{split}
d\sigma_+& =d\sigma_-\propto \frac{1}{2}\left|A_{dp}(1,-\tfrac12,\tfrac12, q)\right|^2\\
& \propto u^2(q) + \frac{1}{2} w^2(q) -\sqrt2 u(q)w(q) ,\\
d\sigma_0& \propto\frac{1}{2}\left(
\left|A_{dp}(0,\tfrac12,\tfrac12, q)\right|^2 +
\left|A_{dp}(0,-\tfrac12,-\tfrac12, q)\right|^2\right)\\
& \propto u^2(q) + 2w^2(q) +2\sqrt2 u(q)w(q),
\end{split}
\ee
and
\be\label{rho20.2}
\rho_{20}=-\frac{1}{\sqrt{2}}\frac{2\sqrt2 u(q)w(q)+w^2(q)}{u^2(q)+w^2(q)}.
\ee

\begin{figure}
  \centering
  \includegraphics[height=0.25\textheight]{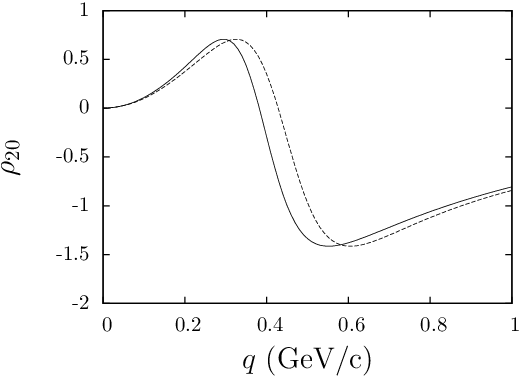}
\caption{Deuteron alignment calculated with $^3$He wave functions for the
Paris (solid) and CD-Bonn (dashed) potentials. }
\label{fig:rho20}
\end{figure}

It is clear from Eq.~(\ref{rho20.2}) that, even in the case of the breakup of
an unpolarized $^3$He, the deuteron spectator emitted at $0^{\circ}$ has a
tensor polarization. Note that at small $q$ this results in $\rho_{20}\approx
-2 w(q)/u(q)=-2q^2D_2$. The predictions for $\rho_{20}$ obtained with the
$^3$He wave functions from the Paris and CD-Bonn potentials are shown in
Fig.~\ref{fig:rho20}.
\begin{table*}
\caption{\label{tab:D2}$D_2(3N)$ parameter (in fm$^2$).}
\begin{ruledtabular}
\begin{tabular}{lllll}
  Paris   &  CD-Bonn  & AV18 \cite{Schiavilla}&  Urbana \cite{Schiavilla}&
  Experiment \cite{Sen}\\
\hline
$-0.2387$   &  $-0.2487$  & $-0.27$                 & $-0.23$                    & $-0.259 \pm 0.014$
\end{tabular}
\end{ruledtabular}
\end{table*}

\subsection{Polarization transfer from $^3$He to $\boldsymbol{d}$}

We consider here the case when the quantization axes for the $^3$He and the
deuteron are parallel and both are perpendicular to the deuteron momentum. We define
the coefficient of the vector-to-vector polarization transfer from polarized
$^3$He to deuteron (whose vector polarization is under consideration) as
\be\label{Pol.1}
\begin{split}
& \kappa_d=\sum_\xi \left[ d\sigma(1,\xi,\tfrac12) +  d\sigma(-1,\xi,-\tfrac12)\right.\\
&\left.-d\sigma(1,\xi,-\tfrac12) -  d\sigma(-1,\xi,\tfrac12)\right] /\sum_{M,\xi,\sigma} d\sigma(M,\xi,\sigma) ,
\end{split}
 \ee
where $d\sigma(M,\xi,\sigma)\propto |A_{dp}(M,\xi,\sigma,q)|^2$.

From invariance under space inversion we have
\be\label{Pol.2}
\begin{split}
& \sum_\xi d\sigma(1,\xi,\tfrac12)=\sum_\xi d\sigma(-1,\xi,-\tfrac12),\\
& \sum_\xi d\sigma(-1,\xi,\tfrac12)=\sum_\xi d\sigma(1,\xi,-\tfrac12),\\
& \sum_\xi d\sigma(0,\xi,\tfrac12)=\sum_\xi d\sigma(0,\xi,-\tfrac12)
\end{split}
\ee
and Eq.~(\ref{Pol.1}) reduces to
\be\label{Pol.3}
\kappa_d=\frac{\sum_\xi \left[ d\sigma(1,\xi,\tfrac12) -d\sigma(1,\xi,-\tfrac12) \right] }
{\sum_{M,\xi,\sigma} d\sigma(M,\xi,\tfrac12) }.
\ee
It is straightforward to show that
\be\label{Pol.4}
\begin{split}
& d\sigma(1,\tfrac12,\tfrac12)=d\sigma(-1,\tfrac12,\tfrac12)=d\sigma(0,-\tfrac12,\tfrac12)=0,\\
&d\sigma(1,-\tfrac12,\tfrac12)\propto \left(-\sqrt{\tfrac23} u +\frac12\sqrt{\tfrac13} w\right)^2 ,\\
& d\sigma(-1,-\tfrac12,\tfrac12)\propto \tfrac34w^2 ,\\
&
d\sigma(0,\tfrac12,\tfrac12)\propto\left(-\sqrt{\tfrac13}u+\tfrac12\sqrt{\tfrac23}w
\right)^2
\end{split}
\ee
and the polarization transfer coefficient becomes
\be\label{Pol.5}
\kappa_d=\frac23\left(\frac{u^2-w^2-uw/\sqrt2}{u^2+w^2}\right)\!. \ee

We point out here that the expression given in Eq.~(\ref{Pol.5}) differs from
Eq.~(5) of Ref.~\cite{sitdeuteron93}  by a factor 2, which
was erroneously lost in that paper.

\begin{figure}
  \centering
  \includegraphics[height=0.25\textheight]{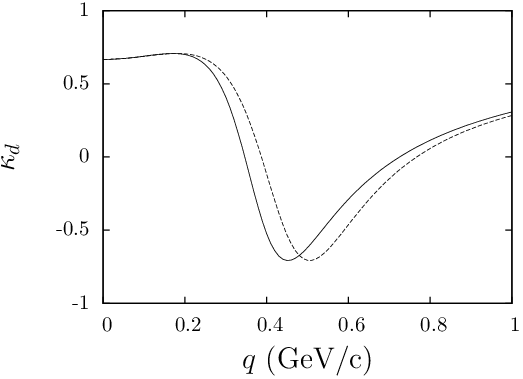}
\caption{Polarization transfer $\kappa_d$ from $^3$He to $d$ for the
Paris (solid) and CD-Bonn (dashed) potentials.}
\label{fig:kappa}
\end{figure}

Results of calculations with $^3$He wave functions from the Paris and CD-Bonn
potentials are shown in Fig.~\ref{fig:kappa}.

The observables $\kappa_d$ and $\rho_{20}$ are related by:
\begin{equation}\label{eq:rho20andkappad}
    \left(\frac{3}{2}\kappa_d\right)^2 + \left(\rho_{20} + \frac{1}{2\sqrt{2}}\right)^2
= \frac{9}{8}.
\end{equation}

Furthermore, at small $q$
\begin{equation}\label{eq:kappadsmallq}
   \kappa_d\approx
    \frac{2}{3}\left(1-\frac{q^2D_2}{\sqrt{2}}  \right)\approx
\frac{2}{3}\left(1+\frac{\rho_{20}}{2\sqrt{2}}  \right),
\end{equation}
so that $\kappa_d\to 2/3$ when $q\to 0$.

\subsection{Polarization transfer from $^3$He to $\boldsymbol{p}$}
The polarization transfer from $^3$He to $p$ is defined by
\be\label{He-p.1}
\kappa_p=\frac{p_{\frac12\frac12}-p_{\frac12-\frac12}}{p_{\frac12\frac12}+p_{\frac12-\frac12}},
\ee
where the $p_{\sigma\xi}$ are given by Eq.~(\ref{p_1212}). At
$\theta=90^\circ$ this reduces to
\be\label{He-p.2}
\kappa_p=\frac{\rho_1-\rho_2-\rho_4-2(\rho_3+\rho_5)+2\sqrt2(\rho_{13}+\rho_{45})}{\rho_{1}+3(\rho_{2}+\rho_{3}+\rho_{4}+\rho_{5})}.
\ee
For the $d+p$ projection of the $^3$He wave function, the
proton momentum distributions are
\be\label{He-p.3}
\begin{split}
\widetilde p_{\frac12\frac12}(q,90^\circ)&=\frac{2\pi^2}{3}\left[ u^2(q) - \sqrt2 u(q)w(q) + \tfrac12 w^2(q)\right] ,\\
\widetilde p_{\frac12-\frac12}(q,90^\circ)&=\frac{2\pi^2}{3}\left[ 2u^2(q) + \sqrt2 u(q)w(q) + \tfrac52 w^2(q)\right]
\end{split}
\ee
and hence
\be\label{He-p.4}
\widetilde \kappa_p=-\frac{u^2+2\sqrt2 uw +2 w^2}{3(u^2+w^2)}.
\ee
\begin{figure}
  \centering
  \includegraphics[width=0.45\textwidth]{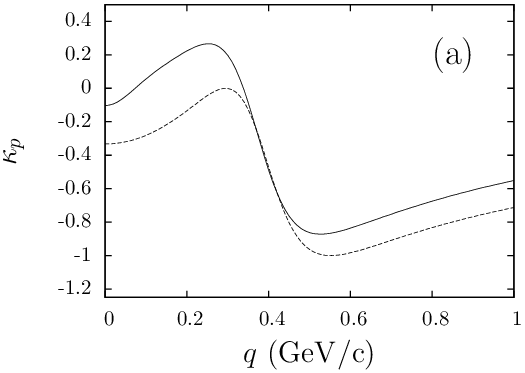}
  \includegraphics[width=0.45\textwidth]{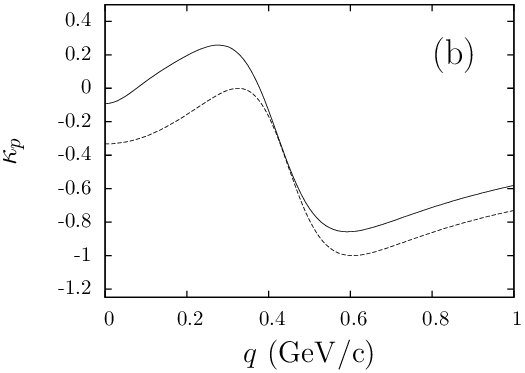}

\caption{The coefficient of polarization transfer from $^3$He to the
proton. The $^3$He wave function used is based on the Paris potential (a) and the
CD-Bonn potential (b). Solid line: full wave function. Short-dashed line:
only the $d+p$ projection (i.e., the $\widetilde \kappa_p$).}
\label{fig:kappastoproton}
\end{figure}
The behavior of $\kappa_p$ and $\widetilde \kappa_p$ is shown in
Fig.~\ref{fig:kappastoproton}.

It is easy to see that the observables $\widetilde\kappa_p$ and $\rho_{20}$
must be related because the spin-dependent observables under consideration
are determined by the ratio of the two functions $u(q)$ and $w(q)$. One then
finds
\begin{equation}\label{eq:pynrho20}
    \widetilde\kappa_p = -\frac{1}{3}\left(1-\sqrt{2} \rho_{20} \right),
\end{equation}
so that at small $q$
\begin{equation}\label{eq:pynrho20smallk}
    \widetilde\kappa_p\approx
    -\frac{1}{3}\left(1- 2\sqrt{2}q^2D_2 \right)
\end{equation}
and hence $\widetilde\kappa_p \to -1/3$ when $q\rightarrow 0$.

A linear combination of the two polarization transfer coefficients has the
following behavior at small $q$:
\begin{equation}\label{eq:kappasrelat}
1-\left(\widetilde\kappa_p+2\kappa_d\right)\approx
3 q^4 (D_2)^2\approx\frac{3}{4}(\rho_{20})^2\  {.}
\end{equation}

As a final remark in this section, we point out that the coefficient of
polarization transfer from $^3$He to the neutron, $\kappa_n$, is equal to
1 in the spectator model. This follows immediately from
Eq.~(\ref{NeuteronMD}).

\section{\label{Sec:Comparing}Comparison with experiment}
\subsection{``Minimal relativization scheme''}
To compare the calculated momentum distributions with experiment, it
is necessary to establish a connection between the argument $\mathbf{q}$ of
the $^3$He wave function and the measured spectator momentum. This must be
done in a way that allows one to take into account relativistic effects in
$^3$He.

This connection can be made in the framework of the so-called ``minimal
relativization scheme.'' In this approach, the relative momentum $\mathbf{q}$
between the pair and the spectator nucleon, taken in the $^3$He rest frame,
is replaced by the relativistic internal momentum
$\mathbf{k}=(\mathbf{k}_\perp,k_l)$, which appears in the light-cone
dynamics. This is also referred to as the dynamics in the infinite momentum
frame (IMF). The IMF is the limiting reference frame which is moving, with
respect to the laboratory frame, in the negative $z$ direction with a
velocity that is close to the speed of light. The important question is how
the light-cone variable $\mathbf k$ is connected with the measured momentum
of the $^3$He fragment.

In the IMF the nucleon momenta are parametrized as
\be\label{IMF_parameterization}
\begin{split}
&p_1=\left((1-\alpha)\mathcal P+\frac{m^2+\mathbf q_\perp^2}{4(1-\alpha)\mathcal P}\,,\ \ \mathbf q_\perp,\right.\\
&\left. \hspace{4.cm}(1-\alpha)\mathcal P-\frac{m^2+\mathbf q_\perp^2}{4(1-\alpha)\mathcal P}\right),\\
&p_2=\left(\beta\mathcal P+\frac{m^2+\left(\mathbf p_\perp-\tfrac12\mathbf q_\perp\right)^2}{4\beta\mathcal P}\,,\ \ \mathbf p_\perp-\tfrac12\mathbf q_\perp,\right.\\
&\left. \hspace{3.5cm} \beta\mathcal P-\frac{m^2+\left(\mathbf p_\perp-\tfrac12\mathbf q_\perp\right)^2}{4\beta\mathcal P}\right),\\
&p_3=\left((\alpha-\beta)\mathcal P+\frac{m^2+\left(\mathbf p_\perp+\tfrac12\mathbf q_\perp\right)^2}{4(\alpha-\beta)\mathcal P}\,,\ -\mathbf p_\perp-\tfrac12\mathbf q_\perp,\right.\\
&\left. \hspace{2.5cm}  (\alpha-\beta)\mathcal P-\frac{m^2+\left(\mathbf p_\perp+\tfrac12\mathbf q_\perp\right)^2}{4(\alpha-\beta)\mathcal P}\right)
\end{split}
\ee
with $0 \leq\alpha\leq 1$ and
$0 \leq\beta\leq \alpha$. Here $\mathcal P\gg 3m$
is the IMF momentum of
$^3$He. The square of the {\it 3N} effective mass is
\be\label{3N_eff_mass}
\begin{split}
(\mathcal M_\mathrm{3N})^2&=(p_1+p_2+p_3)^2 \\
&=\frac{\alpha m^2 +
(1-\alpha)(\mathcal M_\mathrm{2N})^2 + \mathbf
q_\perp^2}{\alpha(1-\alpha)}\;,
\end{split}
\ee
where $(\mathcal M_\mathrm{2N})^2$ is effective mass squared of the pair:
\be\label{2N_eff_mass} (\mathcal M_\mathrm{2N})^2=(p_2+p_3)^2=
\frac{m^2+\left[\mathbf p_\perp+\left(\frac{\beta}{\alpha}-
\tfrac12\right)\mathbf
q_\perp\right]^2}{\frac{\beta}{\alpha}\left(1-\frac{\beta} {\alpha}\right)}\
{.} \ee
The relative momentum between the relativistic spectator nucleon and the pair becomes
\be\label{relativistic_k}
k=|\mathbf k|=\sqrt{\frac{\lambda(\mathcal M_\mathrm{3N}^2,\mathcal M_\mathrm{2N}^2,m^2)}{4\mathcal M_\mathrm{3N}^2}}\,,
\ee
where $\lambda(a,b,c)=a^2+b^2+c^2-2ab-2ac-2bc$.

It should be noted that if the relative momentum in the pair is
non relativistic
\be\label{nonrel_p}
\frac{\beta}{\alpha}\approx \frac12,
\ee
then the effective mass $\mathcal M_\mathrm{2N}^2\approx 4m^2$. In this case
the problem is reduced to the relative motion of two particles with fixed
masses, $m$ and $M=2m$, having spins $\frac12$ and $j$, respectively.

We have calculated the average relative momentum squared $\langle p^2\rangle$
in different partial waves and have found that $\langle
p^2\rangle<0.08~(\mathrm{GeV}/c)^2$ for the Paris potential and $\langle
p^2\rangle\lesssim 0.1~(\mathrm{GeV}/c)^2$ for the CD-Bonn potential when
$q\lesssim 0.6~\mathrm{GeV}/c$. This justifies the 
nonrelativistic description of the relative motion in the pair and we therefore
put $(\mathcal M_\mathrm{2N})^2\approx 4m^2$.

The IMF variables $\alpha$ and $\mathbf q_\perp$ must now be connected with
the appropriate kinematical variables measured in the $(^3\mathrm{He},p)$ and
$(^3\mathrm{He},d)$ breakup experiments. In the laboratory frame we choose
the $z$ axis along the $^3$He momentum $P_\tau=(E_\tau, \mathbf 0_\perp, P)$.

In the case of the $(^3\mathrm{He},p)$ breakup the IMF variables $\alpha$ and
$\mathbf k_\perp$ are defined by
\be\label{alpha;q_perp_for proton}
1-\alpha=\frac{E_p+p_l}{E_\tau + P},\quad \mathbf k_\perp=\mathbf q_\perp\,,
\ee
where $p=(E_p,\mathbf q_\perp, p_l)$ is the proton four-momentum (its components
are given here in the laboratory frame). For the $(^3\mathrm{He},d)$ reaction
the IMF variables are defined by
\be\label{alpha;q_perp_for deuteron}
\alpha=\frac{E_d+d_l}{E_\tau + P},\quad \mathbf k_\perp=-\mathbf q_\perp\,,
\ee
where $d=(E_d,-\mathbf q_\perp,d_l)$ is the deuteron four-momentum.

In both cases the relativistic internal momentum in $^3$He is
\be\label{longitudial_k_for_proton}\begin{split}
&\mathbf k=(\mathbf k_\perp, k_l)\  {;}\\
&k_l=\pm\sqrt{\frac{\lambda(\mathcal M_\mathrm{3N}^2,M^2,m^2)}{4\mathcal M_\mathrm{3N}^2}-\mathbf{k}_\perp^2},\qquad \mathbf{k}_\perp=\pm\mathbf{q}_\perp.
\end{split}
\ee
The signs ``$-$'' and ``$+$'' are chosen for $\alpha<\alpha_0$ and
$\alpha>\alpha_0$ respectively; if $\alpha=\alpha_0$ then $k_l=0$. Here
$\alpha_0=\sqrt{4m^2 + \bf q_\perp^2}/\left( \sqrt{m^2 + \bf
q_\perp^2}+\sqrt{4m^2 + \bf q_\perp^2}\right) $.

In the framework of the IMF dynamics the effective deuteron number is given by
\be\label{rel_d_number}
\begin{split}
\mathcal N_d^{\mathrm{ef}}&=\int d^3kd(k)\\
&=\int_0^1d\alpha\int d^2k_\perp
\frac{\varepsilon_p(k)\varepsilon_d(k)}{\alpha(1-\alpha)\mathcal
M_\mathrm{3N}}\,d(k),
\end{split}
\ee
where $\varepsilon_p(k)=\sqrt{m^2+k^2}$, $\varepsilon_d(k)=\sqrt{M^2+k^2}$.
The combination
\be\label{rel_d_mom_dist}
d^{\mathrm{rel}}(\alpha,\mathbf k_\perp)=\frac{\varepsilon_p(k)\varepsilon_d(k)}{\alpha(1-\alpha)\mathcal M_\mathrm{3N}}d(k)
\ee
can be considered as the relativized momentum distribution of the deuteron in
$^3$He.

The invariant differential cross-section for the $A(^3\mathrm{He},d)$ breakup is
given by
\be\label{CS(t,d)} E_d\frac{d^3\sigma}{d\mathbf p_d}=f_\mathrm{kin}^{(d)}
\sigma_p d^{\mathrm{rel}}(\alpha, \mathbf k_\perp),
\ee
where
\be f_\mathrm{kin}^{(d)}=\frac{\lambda^{\frac12}(W,m^2,M_A^2)}{(1-\alpha)M_A
P}, \ee
$W$ is the missing mass squared and $M_A$ is the mass of the target nucleus.

Similarly, one can get the cross section for the $A(^3\mathrm{He},p)$ breakup by
\be\label{CS(t,p)} E_p\frac{d^3\sigma}{d\mathbf p_p}=f_\mathrm{kin}^{(p)}
\sigma_d (1-\alpha) p^{\mathrm{rel}}(\alpha, \mathbf k_\perp),
\ee
where
\be f_\mathrm{kin}^{(p)}=\frac{\lambda^{\frac12}(W,M^2,M_A^2)}{2\alpha M_A P}
\ee
and
\be p^{\mathrm{rel}}(\alpha,\mathbf k_\perp)=
\frac{\varepsilon_p(k)\varepsilon_d(k)}{\alpha(1-\alpha)\mathcal
M_\mathrm{3N}}p(k)\ee
is the relativized momentum distribution of the protons. The $\sigma_p$ and
$\sigma_d$ factors appearing in Eqs.~(\ref{CS(t,d)}) and (\ref{CS(t,p)}) play
the roles here of normalization factors. Their physical interpretation is
beyond the scope of the present paper.

\subsection{Empirical momentum distributions}

Equations (\ref{CS(t,d)}) and  (\ref{CS(t,p)}) connect differential cross
sections of the $^3$He breakup reactions with the relativized momentum
distributions. Despite the fact that these equations were derived in the
framework of the impulse approximation, one may expect that momentum
distributions extracted from $^{12}$C($^3$He,$p$) and $^{12}$C($^3$He,$d$)
breakup data, using Eqs.~(\ref{CS(t,d)}) and  (\ref{CS(t,p)}), include
effectively effects beyond the impulse approximation, in particular coming
from some quark structure of $^3$He. We therefore call the extracted momentum
distributions ``empirical momentum distributions'' (EMDs) of the
spectators in $^3$He.

\begin{figure*}[t]
  \centering
  \includegraphics[height=0.35\textheight]{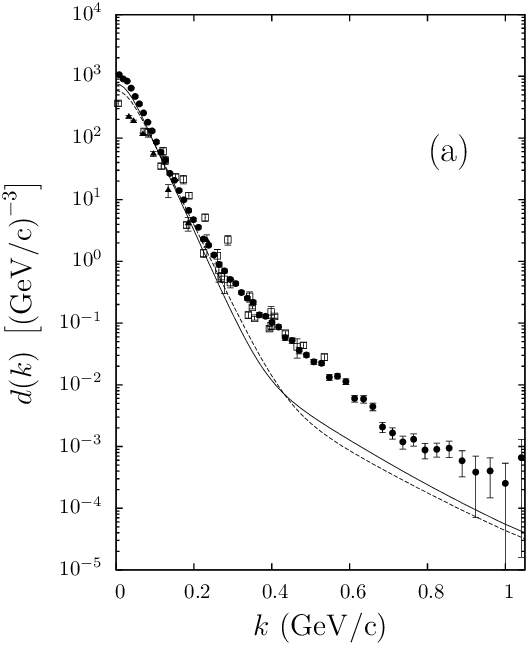}\hspace{1.cm}
  \includegraphics[height=0.35\textheight]{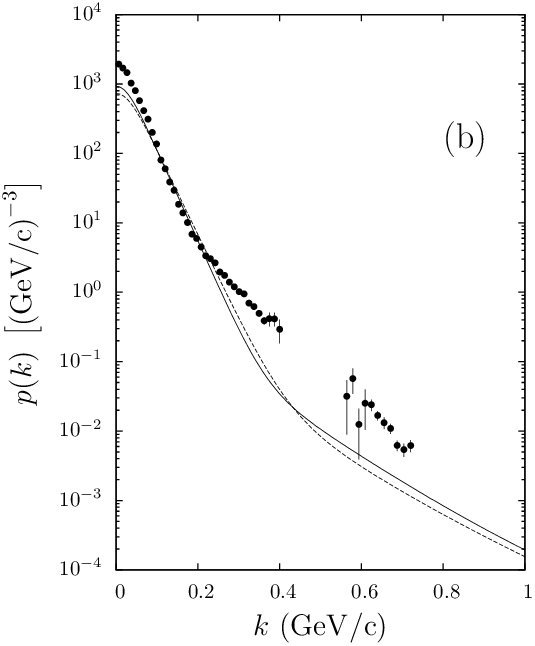}
\caption{
The empirical momentum distribution of the deuterons (a) and the
protons (b) in $^3$He. The solid and dashed lines are calculated with the
Paris and CD-Bonn potentials, respectively. Circles are for the empirical
momentum distributions extracted within Eq.~(\ref{CS(t,d)}) (for deuterons)
and Eq.~(\ref{CS(t,p)}) (for protons) from Ref.~\cite{Ableev3He}. Squares and
triangles represent data extracted from Refs.~\cite{Kitching} and
\cite{Epstein}, respectively. The empirical proton momentum distribution
is normalized to the calculated one for $k<100$~MeV/$c$.
}
\label{fig:experiment}
\end{figure*}

In Fig.~\ref{fig:experiment} we show EMDs for protons and deuterons in $^3$He
extracted from $^{12}$C($^3$He,$p$) and  $^{12}$C($^3$He,$d$) breakup data
obtained for fragments emitted at zero angle and at $p_\mathrm{He}=
10.8$~GeV/$c$~\cite{Ableev3He}.
These EMDs are compared with the results of
our calculations using the Paris and CD-Bonn potentials. The empirical
momentum distribution for the deuteron is also compared with results of other
experiments. Good agreement between data and calculations is obvious at
\underline{small} $k\lesssim 0.25$~GeV/$c$, which indicates that in this
region the spectator model can be used to interpret the data. Note that the
difference between the light cone variable $k$ and the spectator momentum
taken in the $^3$He rest frame is small in this region.

At very small $k\lesssim 50$~MeV/$c$ there is an enhancement of the extracted
EMDs over theoretical curves. A natural explanation of this enhancement
appears to be a
manifestation of Coulomb effects in $^{12}$C($^3$He,$p$) and
$^{12}$C($^3$He,$d$) with registration of the spectator particle at zero
angle~\footnote{~See, e.g., Ref.~\cite{kobushemet}, where a similar
enhancement in the EMD of protons in the deuteron was explained as an effect of
the Coulomb interaction on the mechanism of $^{12}$C($d,p$) breakup.}. We
neglect such Coulomb effects because they take place only over a very narrow
$k$ region and do not affect the data interpretation elsewhere.
Similarly, we neglect any possible final state interaction between the outgoing
protons and deuterons.

It was shown in Ref.~\cite{Ableev3He} that at higher spectator momenta the
EMDs of deuterons and protons in $^3$He plotted versus the relativized
internal momentum $k$ coincide well [which is consistent with
the presentations Figs.~\ref{fig:neutron}(b) and \ref{fig:deutMD}],
while they are significantly different when plotted
versus the 
nonrelativistic internal momentum (the spectator momentum in the
$^3$He rest frame). Such agreement only holds when the analysis is done in
terms of the relativized internal momentum $k$. This indicates that the
$k$-variable is an adequate measure for the internal relative momentum of the
$^3$He constituents.

Data on $(d,p)$ fragmentation~\cite{JINR(dp)}, including those for
spin-dependent observables~\cite{saclay, dubnacarb} and their analysis,
resulted in similar conclusions. At small $k\lesssim 0.25$~GeV/$c$ the
spectator model can be used for the data analysis, the EMD of protons in
the deuteron agrees well with the proton momentum distribution density calculated
for known versions of the deuteron wave function, while at higher $k$ there
is a strong qualitative disagreement between calculations and the existing
set of data. We therefore expect that in the $^3$He breakup for $k\lesssim
250$~MeV/$c$ the reliability of the spectator model predictions should be
about the same as in the $(d,p)$ case.

The data points for momenta above $k\sim 0.25$~GeV/$c$, where the distances
between the $^3$He constituents become comparable to the nucleon radius or
even less, systematically exceed the calculated momentum distributions for
both deuterons and protons. This is once again very similar to the excess of
data over calculations in $(d,p)$ fragmentation~\cite{JINR(dp)}. It was
interpreted as a manifestation of the Pauli principle at the level of
constituent quarks in the two-nucleon system~\cite{Kobushkin}. It is possible
that the observed enhancements in $(^3\mathrm{He},d)$ and $(^3\mathrm{He},p)$
reactions have the same nature. In other words, an extrapolation to this
region of the wave function based on phenomenological realistic {\it NN} potentials
for point-like nucleons is hard to justify.
\subsection{Tensor polarization of the deuteron}
\begin{figure}[b]
  \centering
  \includegraphics[height=0.25\textheight]{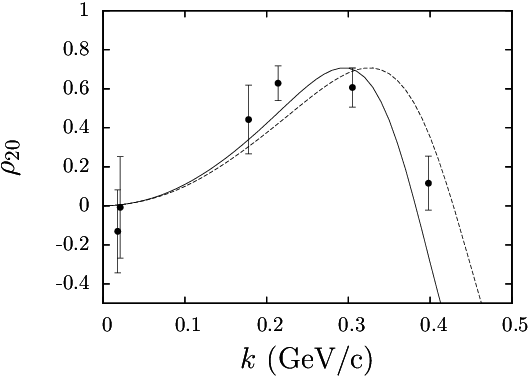}
\caption{Deuteron alignment $\rho_{20}$ calculated with the $^3$He wave
functions for the Paris (solid) and CD-Bonn (dashed) potentials compared with
experimental data. The signs of the data points~\cite{Sitnik} are reversed
to bring them into accordance with the preliminary
results~\cite{Sitnik_prel} of the same experiment, as well as with the sign
of experimental data on the $D_2$ parameter for $^3$He. } \label{fig:rho20_dat}
\end{figure}

Data on the tensor polarization $\rho_{20}$ of the deuteron in the reaction
$^{12}\mathrm{C}(^3\mathrm{He},d)$ at several GeV have been
published~\cite{Sitnik, Sitnik_prel}. It should, however, be noted that the
preliminary data~\cite{Sitnik_prel} extracted from this experiment have the
opposite sign to those tabulated in the final data set~\cite{Sitnik}.

On the other hand, the experimental value of the $D_2$ parameter for $^3$He
projected onto the $d+p$ channel has the opposite sign with respect to the
experimental data on the similar $D_2$ parameter for the deuteron in the
$n+p$ channel. Therefore the sign of the $\rho_{20}$ under discussion must be
opposite to that of the tensor analyzing power in the $d\rightarrow p$
breakup. Taking this into account, together with the contradiction in signs
of $\rho_{20}$ between Refs.~\cite{Sitnik_prel} and \cite{Sitnik}, it is
tempting to conclude that the data tabulated in Ref.~\cite{Sitnik} have the
wrong sign. We therefore use the data from Ref.~\cite{Sitnik} but with a
reversed sign and compare them in Fig.~\ref{fig:rho20_dat} with $\rho_{20}$
calculated according Eq.~(\ref{rho20.2}).

Our results for other spin-dependent observables in the $^3$He breakup cannot
currently be compared with experiment because the relevant data base for
spin-dependent observables is very scarce: at the present time there are no
polarized $^3$He beams with energies of several GeV/nucleon.

\section{\label{Sec:Conclusions}Conclusions}

Using a recent parametrization~\cite{Baru} of the fully antisymmetric
three-nucleon wave function, based on the Paris and CD-Bonn potentials, we
have presented here an analysis of the spin-dependent observables for
$(^3\mathrm{He},d)$ and $(^3\mathrm{He},p)$ reactions, paying special
attention to their behavior at small internal momenta of the $^3$He
fragments. This dependence is determined by a single parameter, known in
the literature as the $D_2$ parameter. Some direct relations between
experimental observables at small internal momenta of the $^3$He fragments
were obtained from this fact.

These relations are useful for the data base cross-checks; at low internal
momenta of the fragments of $^3$He, where the model used for the breakup
reactions works reasonably well, they are rather strict. The $D_2$ parameter
completely determines both the sign and the $q$-dependence of the observables
at small ($\lesssim 150$~MeV/$c$) internal momenta.

On the other hand, these relations show that the breakup reactions with the
lightest nuclei at intermediate energies provide a new way of
obtaining experimental data on the $D_2$ parameter for these nuclei. This is
complementary to the usual methods that involve rearrangement reactions at
low energies.

We emphasize that the different conventions regarding the angular momentum
summations for the 3{\it N} system result in different forms for the formulas
connecting spin-dependent observables with the $^3$He wave function
components. Of course, the final numerical results do not depend on the
conventions provided that the calculations are  performed systematically
within one chosen scheme. However the occasional mixing of the schemes leads
unavoidably to erroneous results. Therefore an explicit indication of the
chosen angular momentum summation scheme is important for the
applications.~\protect\footnote{Perhaps the lack of such indication
explains, why the sign of the $D$ wave, parametrized in Ref.~\cite{GW}
on the basis of values tabulated in Ref.~\cite{Schiavilla}, is opposite to that
of the original tables.\\}

Comparing the results of calculations of the deuteron and proton momentum
distributions in the $^3$He nucleus with existing experimental data, we conclude
that the model used for the $^3$He breakup reactions works reason\-ably
 well for $k\lesssim 250$~MeV/$c$  but at higher momenta  the\\
\footnoterule
\noindent
 data and calculations
are in systematic disagreement. This disagreement, i.e., the enhancement of
the experimental momentum distributions over the calculated ones above
$k\sim 0.25$~GeV/$c$ is very similar to the enhancement of data over calculations
observed for the $(d,p)$ fragmentation~\cite{JINR(dp)} at small emission
angles. This was interpreted for the two-nucleon system as a manifestation of the
Pauli principle at the level of constituent quarks~\cite{Kobushkin}. In other
words, an extrapolation to this region of the wave function based on
phenomenological realistic {\it NN} potentials for pointlike
nucleons is questionable even when relativistic effects are taken into account within the
framework of light cone dynamics.

\acknowledgments The authors gratefully acknowledge Yuri Uzikov for fruitful
discussions on various points considered in the present paper. We are indebted
to Colin Wilkin for his invaluable help with the presentation of the work, as
well as for useful discussions of the model considered. The work of A.P.K
was supported by the Research Programm
``Research in Strong Interacting Matter and Hadron Dynamics in Relativistic
Collisions of Hadrons and Nuclei'' of the National Academy of Sciences of Ukraine.


\appendix
\section{\label{sec:App1}The proton polarization in the $\boldsymbol{(^3}\textrm{He}\boldsymbol{,p)}$ breakup reaction}
The momentum distribution of a proton in $^3\mathrm{He}$ is given by\\[-0.5cm]
\be\label{App.II.1}
\begin{split}
p_{\sigma\xi}(q,\theta)=&3 \int d^3p\left[
\left|\chi_\xi^\dag \left<0010\right|\Psi_\sigma(\mathbf p,\mathbf q)\right|^2 \right.\\
&\left.+\sum_{s_3}\left|\chi_\xi^\dag \left<1s_300\right|\Psi_\sigma(\mathbf p,\mathbf q)\right|^2
\right],\\
\text{~}
\end{split}
\ee
where $\sigma$ and $\xi$ are magnetic quantum numbers for $^3\mathrm{He}$
and the proton, respectively, and $\theta$ is the angle between the quantization
axis and the proton momentum $\mathbf q$. Let us consider the angular
integration of the first and second terms in the square brackets of
Eq.~(\ref{App.II.1}):

\begin{widetext}
\be\label{App.II.2}
\begin{split}
&\int d\Omega_p\left|\chi_\xi^\dag \left<0010\right|\Psi_\sigma(\mathbf p,\mathbf q)\right|^2=
\frac{1}{4\pi} \frac{1}{3}\delta_{\xi,\sigma}\psi^2_1(p,q),
\\
&\sum_{s_3}\int d\Omega_p\left|\chi_\xi^\dag \left<1s_300\right|\Psi_\sigma(\mathbf p,\mathbf q)\right|^2
\\
&=4\pi\sum_{s_3} \left| \frac{1}{4\pi}\left\langle 1 \tfrac12
s_3\xi|\tfrac12\sigma\right\rangle \psi_2(p,q) -
\sqrt{\frac{1}{4\pi}}\left\langle 1 \tfrac12 s_3\xi|\tfrac32 K_3\right\rangle
\left\langle\tfrac322K_3L_3 |\tfrac12\sigma\right\rangle Y_{1L_3}(\widehat
q)\psi_3(p,q)
\right|^2\\
&+\sum_M\left|
-\sqrt{\frac{1}{4\pi}}\left\langle 1 \tfrac12 M\xi|\tfrac12\sigma\right\rangle \psi_4(p,q) +
\left\langle 1 \tfrac12 M\xi|\tfrac32 K_3\right\rangle \left\langle\tfrac322K_3L_3 |\tfrac12\sigma\right\rangle
Y_{1L_3}(\widehat q)\psi_5(p,q)
\right|^2
\\
&=\frac{1}{4\pi}\sum_{i=2,4}\sum_{\rho} \left| \left\langle 1 \tfrac12
\rho\xi|\tfrac12\sigma\right\rangle \psi_i(p,q) - \sqrt{4\pi}\left\langle 1
\tfrac12 \rho\xi|\tfrac32 K_3\right\rangle \left\langle\tfrac322K_3L_3
|\tfrac12\sigma\right\rangle Y_{1L_3}(\widehat q)\psi_{i+1}(p,q) \right|^2\
{.}
\end{split}
\ee

After straightforward calculations we arrive at
\be\label{App.II.3}
\begin{split}
&\sum_{s_3}\int d\Omega_p\left|\chi_{\frac12}^\dag \left<1s_300\right|
\Psi_{\frac12}(\mathbf p,\mathbf q)\right|^2=
\frac{1}{4\pi}\left\lbrace \frac{1}{3}\left[\psi_2^2(p,q) + \psi_4^2(p,q)\right]+
\frac{1}{2}\left( \frac{1}{3}+\cos^2\theta\right)\left[ \psi_3^2(p,q)+ \psi_5^2(p,q)\right] \right.\\
&\left.\hspace{3.5cm}
+\sqrt2\left( \frac{1}{3}-\cos^2\theta\right) \left[ \psi_2(p,q)\psi_3(p,q) + \psi_4(p,q)\psi_5(p,q)\right]\right\rbrace ,  \\
&\sum_{s_3}\int d\Omega_p\left|\chi_{-\frac12}^\dag \left<1s_300\right|
\Psi_{\frac12}(\mathbf p,\mathbf q)\right|^2=
\frac{1}{4\pi}\left\lbrace \frac{2}{3}\left[\psi_2^2(p,q) + \psi_4^2(p,q)\right]+
\frac{1}{2}\left( \frac{5}{3}-\cos^2\theta\right)\left[ \psi_3^2(p,q)+ \psi_5^2(p,q)\right] \right.\\
&\left.\hspace{3.5cm}
-\sqrt2\left( \frac{1}{3}-\cos^2\theta\right) \left[ \psi_2(p,q)\psi_3(p,q) +
\psi_4(p,q)\psi_5(p,q)\right]\right\rbrace\  {.}
\end{split}
\ee
\end{widetext}
%
\end{document}